\documentclass[aps,prx,groupedaddress,reprint,superscriptaddress]{revtex4-2}
\usepackage{amsfonts}
\usepackage{amsmath}\usepackage{amssymb}
\usepackage{graphicx}
\usepackage{dcolumn}
\usepackage{bm}
\usepackage{color}
\usepackage[normalem]{ulem}
\usepackage{multirow}
\begin{document}

\title{Modeling liquid water by climbing up Jacob's ladder in density functional theory facilitated by using deep neural network potentials}

\author{Chunyi Zhang}
\affiliation{Department of Physics, Temple University, Philadelphia, Pennsylvania 19122, USA}

\author{Fujie Tang}
\affiliation{Department of Physics, Temple University, Philadelphia, Pennsylvania 19122, USA}

\author{Mohan Chen}
\affiliation{HEDPS, Center for Applied Physics and Technology, College of Engineering, Peking University, Beijing 100871, China
}

\author{Linfeng Zhang}
\affiliation{Program in Applied and Computational Mathematics, Princeton University, Princeton, New Jersey 08544, USA}	

\author{Diana Y. Qiu}
\affiliation{Department of Mechanical Engineering and Materials Science, Yale University, New Haven, Connecticut 06520, USA}

\author{John P. Perdew}
\affiliation{Department of Physics, Temple University, Philadelphia, Pennsylvania 19122, USA}
\affiliation{Department of Chemistry, Temple University, Philadelphia, Pennsylvania 19122, USA}

\author{Michael L. Klein}
\affiliation{Department of Physics, Temple University, Philadelphia, Pennsylvania 19122, USA}
\affiliation{Institute for Computational Molecular Science, Temple University, Philadelphia, Pennsylvania 19122, USA}
\affiliation{Department of Chemistry, Temple University, Philadelphia, Pennsylvania 19122, USA}

\author{Xifan Wu}
\affiliation{Department of Physics, Temple University, Philadelphia, Pennsylvania 19122, USA}


\begin{abstract}
Within the framework of Kohn-Sham density functional theory (DFT), the ability to provide good predictions of water properties by employing a strongly constrained and appropriately normed (SCAN) functional has been extensively demonstrated in recent years. Here, we further advance the modeling of water by building a more accurate model on the fourth rung of Jacob's ladder with the hybrid functional, SCAN0. In particular, we carry out both classical and Feynman path-integral molecular dynamics calculations of water with the SCAN0 functional and the isobaric-isothermal ensemble. In order to generate the equilibrated structure of water, a deep neural network potential is trained from the atomic potential energy surface based on \emph{ab initio} data obtained from SCAN0 DFT calculations.
For the electronic properties of water, a separate deep neural network potential is trained using the Deep Wannier method based on the maximally localized Wannier functions of the equilibrated trajectory at the SCAN0 level. The structural, dynamic, and electric properties of water were analyzed. The hydrogen-bond structures, density, infrared spectra, diffusion coefficients, and dielectric constants of water, in the electronic ground state, are computed using a large simulation box and long simulation time. For the properties involving electronic excitations, we apply the GW approximation within many-body perturbation theory to calculate the quasiparticle density of states and bandgap of water. Compared to the SCAN functional, mixing exact exchange mitigates the self-interaction error in the meta-generalized-gradient approximation and further softens liquid water towards the experimental direction. For most of the water properties, the SCAN0 functional shows a systematic improvement over the SCAN functional.
\end{abstract}


\maketitle

\section{INTRODUCTION}

Water is arguably the most important substance on this planet. The remarkable functionalities of liquid water arise from a series of its unique and mysterious properties. The density anomaly enables water to freeze from the top down, which is essential to sustain the life and natural evolution on earth \cite{franks_water_2007, 10-Bell}; the frequent autoionization processes and the subsequent fast proton transport determine its amphoteric nature that underpins all the acid-base chemical reactions \cite{geissler_autoionization_2001, 18NC-Chen}; the hydrophobic effect in an aqueous environment is believed to be the driving force behind protein folding \cite{camilloni_towards_2016, baldwin_how_2016}. To unravel all these biological and chemical processes, a genuinely predictive model of water holds the key. Not surprisingly, understanding the hydrogen (H)-bond structure of water continues to command intense attention from both experimentalists and theoreticians \cite{04S-Wernet, 04JPCA-Winter, 08L-Soper, 13JCP-Skinner, 14JCP-Rob, 16L-Chen, 16-Gillan, 17PNAS-Chen, 18NC-Chen, 18JPCL-Gaiduk, 96N-Chandler, pettersson_watermost_2016, nilsson_structural_2015}.

Advanced experimental techniques have been extensively applied to the study of water. The arrangement of water molecules can be probed by neutron scattering or X-ray diffraction experiments \cite{08L-Soper, 13JCP-Skinner}; the H-bond dynamics are detectable by infrared (IR) \cite{max_isotope_2009}, sum-frequency vibrational \cite{shen_sum-frequency_2006, tang_molecular_2020, yang_stabilization_2020}, and pump-probe \cite{woutersen_femtosecond_1997} spectroscopy measurements; the photoemission \cite{04JPCA-Winter} and X-ray absorption \cite{04S-Wernet} spectra provide the electronic structure information of water. Nevertheless, the available experimental techniques only provide ensemble averaged information on water. For insight into the microscopic details, scientists in the field mainly rely on the theoretical modeling of molecules to further understand the structure, dynamics, and electronic properties of water. In order to rationalize the experimental data, an accurate first-principles theory is crucial because its predictions are based on quantum mechanical principles without any empirical input. In this regard, \emph{ab initio} molecular dynamics (AIMD) \cite{85L-CPMD} provides an ideal framework to model liquid water at the microscopic level. In AIMD, the driving force at each timestep is determined by density functional theory (DFT) \cite{hohenberg64, kohn65}.

In DFT, the interacting many-electron problem in any condensed system can be exactly mapped onto a single-body problem with the inclusion of an exchange-correlation (XC) term.
However, the predictive power of DFT crucially depends on the practical form of the XC functional approximation. The accuracy can be systematically improved by climbing up Perdew's metaphorical Jacob's ladder \cite{perdew_jacobs_2001} of XC functionals, with each rung considering additional delicate physical effects. Climbing up Jacob's ladder, which consists of five rungs, to study water is a nontrivial task as more accurate XC functionals require significantly more computational resources. Water is a collection of molecules that are loosely bonded by positively charged H atoms pointing towards negatively charged oxygen (O) atoms, thereby forming the H-bonds that govern many peculiar properties of water. Due to the delicate nature of the H-bond network, it has been widely accepted that the accurate prediction of water properties demands higher rungs of Jacob's ladder than ordinary materials \cite{16-Gillan}. The first rung of Jacob's ladder is the local density approximation (LDA) \cite{80L-Ceperley, 81B-Perdew}, which is parameterized based on the uniform electron gas and tends to minimize the inhomogeneity of electron density. While the overestimation of chemical bonds by LDA was observed as a general trend \cite{laasonen_water_1992, laasonen_structures_1993}, the significantly overestimated H-bond strength means that LDA fails in even qualitatively predicting water properties. Also, first-principles MD calculations of water by LDA only lasted for a short time period in the 1990s. In the following twenty years, simulations of water were dominated by utilizing the second rung of XC functionals with the generalized gradient approximation (GGA) \cite{88A-Becke, 88B-Lee}. By introducing the electron density gradient, the GGA functional allows more constraints to be satisfied than LDA. In particular, the nonempirical GGA functional by Perdew, Burke, and Ernzerhof (PBE) \cite{96L-PBE} satisfies both the constraints on the electron-correlation hole and electron-correlation energy. PBE-GGA weakens the H-bond strengths towards the experimental direction \cite{16-Gillan}. Revisions of PBE such as revPBE \cite{zhang_comment_1998} are constructed by satisfying fewer physical constraints than PBE. However, GGA also has limitations. PBE-GGA intrinsically lacks the intermediate- and long-range van der Waals (vdW) interactions and largely underestimates the liquid water density; notably, incorrectly yielding denser ice than water \cite{11JCP-Wang}. Recently, the non-empirical meta-GGA strongly constrained and appropriately normed (SCAN) functional \cite{15L-Sun, 16NC-Sun} was proposed as a general-purpose \emph{ab initio} functional that belongs to the third rung of Jacob's ladder. With the introduction of kinetic energy density together with a dimensionless variable, SCAN satisfies 17 known constraints on XC functionals as well as the tight lower bound on the exchange energy which has been elusive in PBE-GGA on the second rung of Jacob's ladder. SCAN was found to provide a substantially more accurate description of water as compared to GGAs, including structural, electronic, and dynamical properties \cite{17PNAS-Chen, 18JCP-Zheng, xu_importance_2021}. Most importantly, SCAN includes intermediate-range vdW interactions and the resulting density of water is 1.050 $g/cm^3$, which is much closer to the experimental value of 0.997 $g/cm^3$, as compared to the density of $\sim$ 0.850 $g/cm^3$ from PBE-GGA functionals \cite{17PNAS-Chen}. In the past few years, SCAN has been one of the most commonly used non-empirical XC functionals in the studies of water.

Notwithstanding the above significant progress, challenges remain. Both GGA and meta-GGA functionals suffer from the self-interaction error \cite{perdew_self-interaction_1981, sharkas_self-interaction_2020}. The spurious interaction of the electron with itself leads to excessive proton delocalization and an artificially strengthened H-bond \cite{zhang_first_2011}. The overestimated H-bond strength is evidenced by the red-shift of the IR stretching band predicted by SCAN in comparison with experiment \cite{xu_first-principles_2019}. The above is also consistent with the slightly overstructured oxygen-oxygen pair distribution predicted by SCAN as compared to the neutron scattering data \cite{17PNAS-Chen}. Furthermore, most AIMD simulations of water so far have assumed classical nuclei.  However, the light hydrogen atom deviates from classical behavior significantly, even at room temperature  \cite{li_quantum_2011, ceriotti_nuclear_2013, ceriotti_nuclear_2016}. The Feynman discretized path-integral (PI) method combined with AIMD can treat the nuclear quantum effect (NQEs) \cite{marx_ab_1996, ceriotti_i-pi:_2014}. But for reasons of computational cost, the PI-AIMD simulations have not yet been widely applied to water. More disturbingly, many thermodynamic properties of water are difficult to predict by DFT based on currently available computational resources. Standard AIMD simulations are typically carried out for a time scale of tens of picoseconds, and the system size is limited to a simulation box containing a few tens to a hundred or so water molecules.
In water, the time for the H-bond to break and reform is typically at the picosecond scale \cite{fecko_ultrafast_2003}. Therefore, the H-bond structure in water can be faithfully predicted within tens of picoseconds by AIMD simulation. However, for many macroscopic properties, such as diffusivity and dielectric constant, the converged predictions require much longer simulation time and larger system size. In order to approach the thermodynamic limit, at least thousands of water molecules are necessary to be considered in the simulation box, which is simply impossible for AIMD.

To address the above issues, we ascend Jacob's ladder of DFT to the fourth rung, using the SCAN0 hybrid functional to study liquid water with NQEs considered simultaneously by the PI method. The predictions of converged thermodynamic properties are largely facilitated by employing state-of-the-art deep potential molecular dynamics (DPMD) \cite{18L-Zhang, zhang_active_2019, zhang2018end}. The many-body potential and interatomic forces needed for DPMD are obtained from a deep neural network that is trained with SCAN0 DFT data.
DPMD has recently been demonstrated to be able to accurately model the structure of pure water with an accuracy at the DFT level and with the low computational cost of a classical force field \cite{18L-Zhang, ko_isotope_2019}. For the electronic properties, we adopt the Deep Wannier \cite{zhang_deep_2020} neural network model, which is trained with the positions of centers of the maximally localized Wannier functions \cite{97MLWF, 12MLWF} that are obtained from SCAN0 DFT calculations. Compared to the third rung of Jacob's ladder of SCAN, SCAN0 presents a better description of liquid water in terms of structure, dynamics, as well as electronic properties. By mixing 10\% exact change as suggested for an earlier meta-GGA functional \cite{03JCP-Staroverov}, SCAN0 mitigates the self-interaction error. The less polarizable water together with the NQEs on the protons yield a liquid structure softened towards the experimental direction compared with SCAN. The predicted dynamic properties such as IR spectra and diffusivities and the electric properties such as dielectric constants are in good agreement with available experiments. Finally, we compute the quasiparticle band structure of liquid water based on Hedin's GW approximation  \cite{hedin_new_1965} as implemented in the BerkeleyGW package \cite{hybertsen_electron_1986, deslippe_berkeleygw_2012}. The predicted density of states (DOS) and bandgap are in reasonable agreement with recent photoemission spectroscopy (PES) experiments.

\section{Methods}

\subsection{The training process of the SCAN0 deep potential model}
To construct an accurate and transferable SCAN0 path-integral deep potential model with a minimal number of the expensive SCAN0 DFT data, an active machine learning procedure called deep potential generator (DP-GEN) \cite{zhang_active_2019} was adopted. The procedure is summarized as follows:

(1) Considering that the molecular configurations predicted by SCAN and SCAN0 are close to each other, 900 configurations were uniformly extracted from the 64-molecule 11 ps SCAN PI-AIMD trajectory reported in Ref.~\cite{zhang_isotope_2020}. The total potential energy $E$ and ionic forces {$\textbf{\emph{F}}_i$} of each atom $i$ of these configurations were calculated using Quantum ESPRESSO (QE) \cite{17CPCM-QE} with the SCAN0 XC functional. The Hamann-Schl\"{u}ter-Chiang-Vanderbilt (HSCV) pseudopotentials \cite{79L-Hamann,85B-Vanderbilt} with an energy cutoff of 150 Ry were employed.
The obtained $E$ and {$\textbf{\emph{F}}_i$} together with atomic positions were adopted as the initial training data.

(2) The DeePMD-kit package \cite{wang_deepmd-kit:_2018} was used to train four deep potential models independently for $10^{6}$ steps. The differences between the training processes of the four deep potential models were the initialization random parameters and the activation function. Two models were trained using the Tanh activation function, while the other two models were trained using the GELU activation function.
The training followed the procedure described in Refs.~\cite{18L-Zhang, zhang2018end}.
First, the input data were transformed to local coordinate frames for every atom and its neighbors inside a cutoff distance of 6 \r{A} to preserve the translational, rotational, and permutational symmetries. Afterward, the deep neural network parameters were optimized by the Adam method \cite{kingma_adam_2017} with the loss function:
$$\mathcal{L}(p_\epsilon,p_f)=p_{\epsilon}\Delta\epsilon^2+\frac{p_f}{3n}\sum_i|\Delta{\textbf{\emph{F}}_i}|^2,$$
where $\Delta\epsilon$ and ${\Delta}\textbf{\emph{F}}_i$ represent the differences between the training data and current deep potential prediction for the quantities $\epsilon\equiv{E}/n$ and \textbf{\emph{F}}$_i$, respectively, $n$ is the number of atoms, $p_\epsilon$ and $p_f$ are tunable prefactors. In the training process, $p_\epsilon$ progressively increases from 0.02 to 1, while $p_f$ progressively decreases from 1000 to 1.

(3) To explore the potential energy surface, ten independent PI-DPMD simulations were carried out in the isobaric-isothermal (\emph{NpT}) ensemble at 1 bar and 330 K for 100 ps with a timestep of 0.5 fs using the i-PI code \cite{ceriotti_i-pi:_2014} in conjunction with the DeePMD-kit package \cite{wang_deepmd-kit:_2018}. A periodically replicated cubic cell containing 64 water molecules was used. The nuclear degrees of freedom were sampled using eight beads with a colored-noise generalized Langevin equation thermostat \cite{ceriotti_nuclear_2009, ceriotti_efficient_2012} to accelerate the convergence of the quantum distribution.
The pressure was controlled using an isotropic barostat as implemented in the i-PI code with a time constant associated with the dynamics of the piston of 200 fs.
The cell was thermostatted by another generalized Langevin equation thermostat \cite{ceriotti_langevin_2009, ceriotti_colored-noise_2010} to enhance sampling efficiency.
For each PI-DPMD simulation, all of the four deep potential models were used to predict the force acting on each atom at each timestep but only one model randomly selected from the four models was used to propagate the trajectory.

(4) The convergence of the PI-DPMD simulations was checked. Following Ref.~\cite{zhang_active_2019}, the maximum standard deviation of the predicted atomic forces $\zeta=max_{i}\sqrt{{\langle}\|\textbf{\emph{f}}_i-\overline{\textbf{\emph{f}}}_i\|{\rangle}}$ was adopted as an indicator for the convergence, where $\overline{\textbf{\emph{f}}}_i={\langle}\textbf{\emph{f}}_i{\rangle}$ is the average force on atom $i$ predicted by the four different deep potential models.
If the input training data set can cover the potential energy surface, the four forces predicted by the four models at each PI-DPMD step will agree with each other and $\zeta$ will close to zero. In this work, when the configurations that have $\zeta >$ 0.2 eV/\r{A} accounts for less than 0.005 \% of all the PI-DPMD configurations, the models were considered to be converged. Otherwise, 50-400 configurations that have $\zeta >$ 0.2 eV/\r{A} were extracted. The force and energy of these configurations were calculated using QE with the SCNA0 using the same DFT parameters as step one. The new DFT results were added to the training data set and the loop was repeated until obtaining a converged deep potential model.

\begin{figure}
	\includegraphics[width=0.4\textwidth]{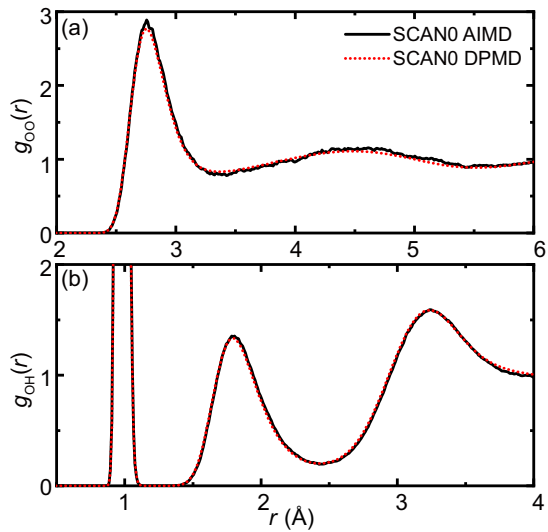}
	\caption{(a) ${g\rm_{OO}}(r)$ and (b) ${g\rm_{OH}}(r)$ of liquid H$_2$O from SCAN0 AIMD and DPMD simulations at 1 bar and 330 K.  }\label{fig:convergence}
\end{figure}

The deep potential model used in the production run was trained using 7349 SCAN0 DFT configurations. To verify the reliability of the deep potential model, we also conducted a SCAN0 AIMD simulation of liquid water in the $NpT$ ensemble at 1 bar and 330 K for 20 ps with a 64-molecule cell using QE. HSCV pseudopotentials \cite{79L-Hamann,85B-Vanderbilt} with an energy cutoff of 130 Ry were adopted.
As shown in Fig.~\ref{fig:convergence}, the RDFs predicted by AIMD and DPMD simulations agree well with each other, which proved the convergence of the SCAN0 deep potential model.

To predict the electronic properties, the deep Wannier \cite{zhang_deep_2020} method was applied to build another deep potential model to map input atomic coordinates into output Wannier centers.
Firstly, the Wannier centers of the 7349 SCAN0 DFT configurations were calculated using QE. Each water molecule $i$ has four valence Wannier centers. The centroid of the four Wannier centers $\textbf{\emph{r}}_w^i=\frac{1}{4}\sum_{c=1}^{4}\textbf{\emph{r}}_{w_c}^i$ and the atomic positions were adopted as the input training data and trained for $10^6$ steps using the DeePMD-kit package \cite{wang_deepmd-kit:_2018}.

\subsection{Classical and quantum DPMD simulations}

The SCAN0 deep potential model was applied to conduct both DPMD and PI-DPMD simulations at 1 bar and 330 K with a simulation cell containing 512 water molecules. Specially, the dielectric constant and diffusion coefficient were calculated using cells containing 64$\sim$4096 molecules. All static properties were calculated using the $NpT$ ensemble. To avoid the influence of the thermostat and barostat on the dynamics, all dynamical properties were calculated using the microcanonical (\emph{NVE}) ensemble with the cell size fixed at the value obtained from the $NpT$ simulation.
The DPMD simulation was carried out using LAMMPS \cite{plimpton_fast_nodate} in conjunction with the DeePMD-kit package \cite{wang_deepmd-kit:_2018} for 2 ns with the first 100 ps discarded for equilibrium.
The PI-DPMD simulation was conducted using the i-PI code \cite{ceriotti_i-pi:_2014} in connection with the DeePMD-kit package \cite{wang_deepmd-kit:_2018} for 500 ps with the first 100 ps discarded for equilibrium. The PI-DPMD simulation parameters related to nuclear and cell dynamics were the same as those adopted in step three of the deep potential training process.

\subsection{Calculation of the electronic DOS and bandgap}
The electronic DOS and bandgap of liquid water were calculated at the level of DFT and G$_0$W$_0$.
All the DFT calculations reported in this work are carried out at the level of SCAN0. A 0.1 eV Gaussian broadening factor was adopted for the computed eigenstates.
The SCAN0 classical DFT result was calculated based on 200 snapshots that uniformly extracted from the 2 ns 64-molecule SCAN0 DPMD trajectory.
The SCAN0 quantum DFT result was calculated based on 100 snapshots (each snapshot containing 8 beads) that uniformly extracted from the 500 ps 64-molecule SCAN0 PI-DPMD trajectory, both SCAN0 classical and quantum DFT results containing 256 valence bands and 10 conduction bands.

The G$_0$W$_0$ calculation was carried out by including dynamical dielectric screening effects using a generalized plasmon-pole model with the BerkeleyGW package \cite{hybertsen_electron_1986, deslippe_berkeleygw_2012}. Due to the large computational cost of the G$_0$W$_0$ method, the SCAN0 quantum result was calculated using 8 snapshots 
uniformly extracted from the 500 ps 32-molecule SCAN0 quantum trajectory, with the PBE functional as a starting point. For all the configurations, the dielectric function $\epsilon$ and the self-energy $\Sigma$ were calculated using 20000 bands to ensure convergence. The plane wave energy cutoff for $\epsilon$ was 30 Ry. 
A single gamma point was adopted. Overall, 128 occupied eigenstates and 5 empty eigenstates were calculated.

\section{RESULTS AND DISCUSSION}

\subsection{Water structure and density}

\begin{table*}[ptb]
	\caption{The structural, dynamical, and electronic properties of liquid water predicted by SCAN0 DPMD and PI-DPMD simulations at 330 K in comparison with the 330 K SCAN AIMD results and the room temperature experimental results. The structural properties include the average O-H covalent bond length ($d_{\rm OH}$), average number of H-bonds per water molecule ($N_{\rm HB}$), and density ($\rho$). The dynamical properties include the peak positions of the four IR bands ($\nu_{\rm T}$, $\nu_{\rm L}$, $\nu_{\rm B}$, $\nu_{\rm S}$), the diffusion coefficient ($D$), and the rotational correlation time ($\tau_2$). The electronic properties include the dipole moment ($\mu$), the binding energies of $2a_1$, $1b_2$, $3a_1$ peaks with respect to the $1b_1$ peak in the electronic DOS, the bandgap, and the dielectric constant ($\epsilon_0$). For the method column in the electronic part, the texts outside/inside the parentheses represent the calculation method of the molecular/electronic structure.
}
	\label{tbl_compare}
    \renewcommand\arraystretch{1.1}
    \begin{ruledtabular}
		\begin{tabular}{llllllll}
        Property &Method  &$d_{\rm OH}$ (\r{A}) &$N_{\rm HB}$ &$\rho$ ($g/cm^3$)   \\
	    \colrule
        \multirow{3}*{Structural} &Exp. &1.01 \cite{08L-Soper}  & &0.997 \cite{linstrom_nist_2001} \\
        &SCAN AIMD &0.984 \cite{17PNAS-Chen}&3.61 \cite{17PNAS-Chen}&1.050 \cite{17PNAS-Chen} \\
        (H$_2$O) &SCAN0 DPMD &0.980 &3.58 &1.030 \\
        &SCAN0 PI-DPMD &0.999 &3.49 &1.041 \\
	    \colrule
              &Method  &$\nu_{\rm T}$ ($cm^{-1}$) &$\nu_{\rm L}$ ($cm^{-1}$) &$\nu_{\rm B}$ ($cm^{-1}$) & $\nu_{\rm S}$ ($cm^{-1}$) & $D$ (\r{A}$^2/ps$) & $\tau_2$ (ps)\\
	    \colrule
        \multirow{3}*{Dynamical} &Exp. &186 \cite{max_isotope_2009}&486 \cite{max_isotope_2009}&1209 \cite{max_isotope_2009} & 2498 \cite{max_isotope_2009}& 0.20 \cite{73JPC-Mills} &2.4 \cite{ropp_rotational_nodate}\\
        &SCAN AIMD  &172 \cite{xu_first-principles_2019}&483  \cite{xu_first-principles_2019}&1207  \cite{xu_first-principles_2019}& 2448  \cite{xu_first-principles_2019}& \\
        (D$_2$O)&SCAN0 DPMD &227 &486 &1219 & 2507 & 0.26 &1.9\\
        \colrule
        \multirow{3}*{Dynamical} &Exp. &176 \cite{max_isotope_2009}&615 \cite{max_isotope_2009}&1646 \cite{max_isotope_2009}&3406 \cite{max_isotope_2009}&0.24 \cite{73JPC-Mills} &1.7$\sim$2.6 \cite{lankhorst_determination_1982, smith_proton_1966}\\
        &SCAN & & & & &0.26 \cite{yao_temperature_2020}&3.3 \cite{yao_temperature_2020}\\
        (H$_2$O)&SCAN0 DPMD &208 &675 &1667 & 3475 & 0.29 &1.7\\
	    \colrule
              &Method & $\mu$ (Debye)  &$\epsilon_{2a_1-1b_1}$ (eV)&$\epsilon_{1b_2-1b_1}$ (eV) &$\epsilon_{3a_1-1b_1}$ (eV) &bandgap (eV) &$\epsilon_0$ \\
	    \colrule
        \multirow{5}*{Electronic} &Exp. & 2.9$\pm$0.6 \cite{00JCP-Badyal} &-19.7  \cite{winter_full_2004}&-6.2  \cite{winter_full_2004}&-2.3 \cite{winter_full_2004} &8.7$\pm$0.6 \cite{bernas_electronic_1997}&78.39 \cite{vidulich_dielectric_nodate} \\
        &SCAN AIMD (DFT) &2.97 \cite{17PNAS-Chen} &-18.9 \cite{17PNAS-Chen}&-5.7 \cite{17PNAS-Chen}&-2.0 \cite{17PNAS-Chen}&  & \\
        &SCAN0 DPMD (DFT) & 2.96  &-19.4 &-5.8 &-2.0 &5.1 &76.06 \\
        (H$_2$O)&SCAN0 PI-DPMD (DFT) & 3.08 &-19.4 &-5.6 &-2.0 &4.6 &83.57 \\
        &SCAN0 PI-DPMD (G$_0$W$_0$)& &-19.4 &-6.4 &-2.6 &7.5&   \\
		\end{tabular}
     \end{ruledtabular}
\end{table*}

\begin{figure*}
	\includegraphics[width=0.95\textwidth]{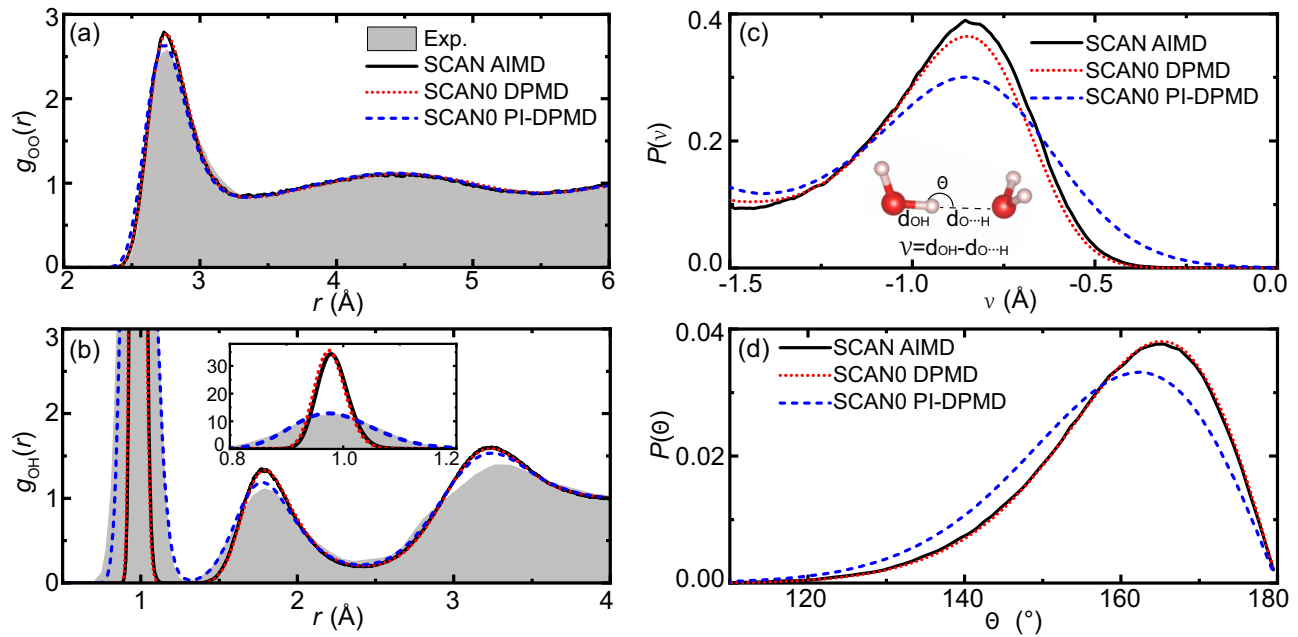}
	\caption{(a) ${g\rm_{OO}}(r)$ and (b) ${g\rm_{OH}}(r)$ of liquid H$_2$O from SCAN AIMD \cite{17PNAS-Chen}, SCAN0 DPMD, and SCAN0 PI-DPMD simulations at 330 K as well as the diffraction experimental result \cite{08L-Soper} at 298 K reported in Refs.~\cite{08L-Soper, zeidler_isotope_2012}. The inset in (b) shows a rounded picture of the first peak of ${g\rm_{OH}}(r)$. The corresponding probability distribution of (c) proton transfer coordinate, $\nu$, and (d) OH$\cdots$O angle, $\theta$ of liquid H$_2$O. The inset in (c) displays definitions of $\nu$ and $\theta$.}\label{fig:gr}
\end{figure*}

We first inspect the structure of water predicted by various DFT functional approximations. To this end, we analyse the radial distribution functions (RDFs), which show the probability to find a given pair of atoms as a function of distance in real space. The resulting oxygen-oxygen and oxygen-hydrogen RDFs, ${g\rm_{OO}}(r)$ and ${g\rm_{OH}}(r)$, are shown in Figs.~\ref{fig:gr} (a) and (b), respectively.
The RDFs predicted by SCAN0 DPMD and SCAN0 PI-DPMD simulations are presented in the same figures with SCAN AIMD \cite{17PNAS-Chen} and experimental results \cite{08L-Soper} for comparison.

Like all local and semi-local DFT functionals, the SCAN functional inherits the spurious self-interaction error, which yields an artificially strengthened tetrahedral structure and delocalized protons. By mixing a fraction (10\%) \cite{03JCP-Staroverov} of exact exchange in SCAN0, the above self-interaction error is mitigated which softens the liquid structure towards the experimental direction. The weakened H-bond strength by SCAN0 DPMD relative to SCAN AIMD is evidenced by the right-shift of the first peak position of ${g\rm_{OO}}(r)$ in Fig.~\ref{fig:gr} (a), from 2.748 to 2.754 \r{A} and the second peak position of ${g\rm_{OH}}(r)$ in Fig.~\ref{fig:gr} (b), from 1.785 to 1.793 \r{A}, both of which are closely associated with the H-bonds. The weaker H-bonds, in turn, enhance the covalency of water molecules in the liquid, as indicated by the shortened OH covalent bond length from 0.984 \r{A} (SCAN AIMD) to 0.980 \r{A} (SCAN0 DPMD). Not surprisingly, the weakened H-bonding also promotes the population of broken H-bonds. Adopting the definition of H-bonds by Chandler \emph{et al.} \cite{96N-Chandler}, the average calculated number of H-bonds per water molecule decreases from 3.61 (SCAN AIMD) to 3.58 (SCAN0 DPMD). The H-bond is highly directional. The forming or breaking of an H-bond critically depends on the geometric configurations as determined by the distance and angle between the H-bond donor and acceptor. As schematically shown in the inset of Fig.~\ref{fig:gr} (c), the proton displacement along the stretching mode towards the H-bond acceptor facilitates the forming of H-bonds; on the other hand, the proton displacement along the libration mode tends to break H-bonds. The above two opposite effects are illustrated by the probability distributions of the proton transfer coordinate, $\nu$, and OH$\cdots$O angle, $\theta$, \cite{wang_quantum_2014, zhang_isotope_2020} as shown in Figs.~\ref{fig:gr} (c) and (d), respectively. It can be seen in Fig.~\ref{fig:gr} (c) that the broken H-bond by the SCAN0 functional originates from the reduced proton displacement along the H-bonding direction. In contrast, from SCAN to SCAN0 functional, the proton displacement along the libration direction barely changes as displayed in Fig.~\ref{fig:gr} (d). This is consistent with the fact that the hybrid functional reduces the electronegativity of the lone pair electrons, which weakens the directional H-bonding strength. In accordance with the reduced directional H-bond strength, the water molecules are more loosely bonded with an increased O-O distance. Therefore, the predicted water density by SCAN0 DPMD is 1.030 $g/cm^3$, which is noticeably smaller than the density of 1.050 $g/cm^3$ predicted by SCAN AIMD simulation.

In SCAN0 PI-DPMD simulations, the NQEs are incorporated in the modeling of water. The resulting ${g\rm_{OO}}(r)$ and ${g\rm_{OH}}(r)$ are presented in Figs.~\ref{fig:gr} (a) and (b), respectively. In quantum simulations, the configuration space explored by protons is largely extended, in particular for the regions that are inaccessible to classical nuclei. Compared to classical simulations, PI-DPMD simulation results in a significantly broadened first peak of ${g\rm_{OH}}(r)$, which is a typical effect reported in all path-integral simulations of water \cite{ceriotti_nuclear_2013, wang_quantum_2014, ceriotti_nuclear_2016, cheng_ab_2019, zhang_isotope_2020, xu_isotope_2020}. Besides the broadening effect, NQEs affect the H-bond network as well. Under the influence of NQEs, the protons are more delocalized along the direction of the stretching mode as well as the direction of the libration mode as shown in Figs.~\ref{fig:gr} (c) and (d). The proton delocalization along the stretching mode yields more polarizable water with a larger electric dipole moment of 3.08 Debye compared to the classical water model of 2.96 Debye, which facilitates the formation of H-bonds. However, the proton delocalization along the libration mode tends to break more H-bonds. In SCAN0 PI-DPMD trajectories, the latter effect is more significant as evidenced by the largely shifted peak position away from H-bond direction (180$^\circ$) in the OH$\cdots$O angle distribution function in Fig.~\ref{fig:gr} (d); in contrast, the center of the proton transfer probability function is less affected by NQEs. As a result of the above two competing NQEs, the average number of H-bonds per water molecule decreases from 3.58 to 3.49. At the same time, the water structure predicted by PI-DPMD is further softened towards the experimental direction as shown in Figs.~\ref{fig:gr} (a) and (b) which is qualitatively similar to results predicted by the path-integral simulations that are based on PBE and BLYP GGA functionals \cite{wang_quantum_2014}. Because of the more polarizable nature of water molecules under NQEs, the H-bonding force along the stretching mode becomes more attractive, which prefers a slightly more compactly packed water structure. Indeed, the predicted water density by SCAN0 PI-DPMD is 1.041 $g/cm^3$, which is slightly larger than 1.030 $g/cm^3$ as predicted by SCAN0 DPMD. The larger density under NQEs has also been widely reported in previous studies \cite{cheng_ab_2019, zhang_isotope_2020, xu_isotope_2020}.

\subsection{Dynamic properties}

The prediction of the quantum dynamic properties of water is still an open problem due to the application of the complex form of the real-time propagator in the current implementations of Feynman's path integral approach \cite{markland_nuclear_2018}. Therefore, the dynamical properties, including the IR spectra and the diffusivity presented in this work, are generated from classical simulations only.

\subsubsection{IR spectrum}
\begin{figure}
	\includegraphics[width=0.45\textwidth]{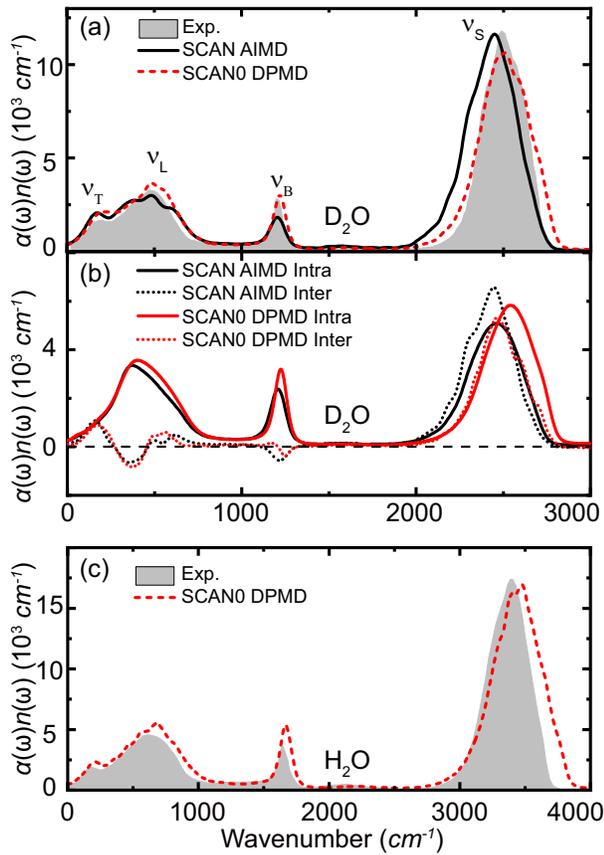}
	\caption{(a) IR spectra of liquid D$_2$O predicted by SCAN AIMD \cite{xu_first-principles_2019} and SCAN0 DPMD simulations at 330 K and the room-temperature experimental result \cite{max_isotope_2009}. (b) Decomposition of the IR spectra in (a) into intra- and inter-molecular contributions. (c) IR spectra of H$_2$O from SCAN0 DPMD simulation at 330 K and room-temperature experimental result \cite{max_isotope_2009}.}\label{fig:ir}
\end{figure}

IR spectroscopy is a powerful experimental technique to probe the vibrational properties of liquid water. The IR spectrum is sensitive to the underlying H-bond network in liquid water and is, therefore, widely adopted to study the microscopic structure of liquid water \cite{08B-Car, zhang_first_2011, xu_first-principles_2019}. Within linear response theory, the IR absorption coefficient per unit length is given by the Fourier transform of the time correlation function of the overall dipole moment as following \cite{ir2000}:
\begin{equation}\label{eq:ir}
\alpha(\omega)=\frac{2\pi\beta\omega^2}{3cVn(\omega)}\int_{-\infty}^{+\infty}dt{e}^{-i\omega{t}}{\langle}\sum_{i,j}\bm{\mu}_i(0)\cdot\bm{\mu}_j(t){\rangle},
\end{equation}
where $n(\omega)$ is the refractive index, $V$ is the volume, and $\beta=(k_B{T})^{-1}$, with $k_B$ and $T$ being Boltzmann's constant and temperature, respectively. The dipole moment $\bm{\mu}_i$ of molecule $i$ is calculated as $\bm{\mu}_i=6\textbf{\emph{r}}_{\rm O}^i+\textbf{\emph{r}}_{{\rm H_1}}^i+\textbf{\emph{r}}_{{\rm H_2}}^i-2\sum_{c=1}^{4}\textbf{\emph{r}}_{w_c}^i$, where $\textbf{\emph{r}}_{\rm O}^i$, $\textbf{\emph{r}}_{{\rm H_1}}^i$, and $\textbf{\emph{r}}_{{\rm H_2}}^i$ is the nuclear coordinates of the $i$th molecule and $\textbf{\emph{r}}_{w_c}^i$ ($c=1{\sim}4$) is the four Wannier centers assigned to the $i$th water molecule in the system, which is obtained from the Deep Wannier method in this work.
Therefore, a precise prediction of the IR spectra demands an accurate description of both the molecular configuration and its dynamical electronic response.

We present the IR spectrum of heavy water predicted by SCAN0 DPMD simulation in Fig.~\ref{fig:ir} (a). For comparison, the spectra computed from the SCAN AIMD simulation \cite{xu_first-principles_2019} and the experimental result \cite{max_isotope_2009} are also presented in the same figure. As shown in Fig.~\ref{fig:ir} (a), the IR spectrum of liquid water displays four main spectral features that can be attributed to the hindered translation ($\nu_{\rm T}$), libration ($\nu_{\rm L}$), H-O-H bending ($\nu_{\rm B}$), and O-H stretching ($\nu_{\rm S}$) modes as a function of increasing frequency. The bending mode and stretching mode can be traced back to the molecular vibration of water vapor in the gas phase, while the hindered translation and libration modes arise from the collective motion of water molecules in the H-bond network. The IR spectra of D$_2$O in Fig.~\ref{fig:ir} (a) show that, compared to SCAN, SCAN0 improves all the four spectral features towards the experimental direction, in terms of both the spectral energies and their intensities. In order to elucidate the physical origin of the improved spectrum, we further decompose the IR spectrum into its intra- and inter-molecular contributions \cite{08B-Car}. The above can be rigorously implemented by separating the time-correlation function in Eq.~\ref{eq:ir} into contributions from $i=j$ and $i{\neq}j$ terms utilizing the additive nature of the two-body correlation function. The resulting decompositions are shown in Fig.~\ref{fig:ir} (b).

We first inspect the stretching band centered around 2500 $cm^{-1}$. According to previous studies, it was found that SCAN functional improves the underestimated stretching frequency compared to the PBE functional \cite{xu_first-principles_2019}. However, the predicted center of the stretching band at $\nu_{\rm S}$=2448 $cm^{-1}$ \cite{xu_first-principles_2019} is still 50 $cm^{-1}$ smaller than the experimental value of 2498 $cm^{-1}$ \cite{max_isotope_2009}, which indicates that the H-bond strength is still overestimated by SCAN functional. Here, the above overestimated H-bond strength is largely corrected by the SCAN0 functional. The SCAN0 functional yields a stretching band centered at $\nu_{\rm S}$=2507 $cm^{-1}$, which is in quantitative agreement with the experimental measurement. We attribute the improvement of the IR stretching band to the reduced self-interaction error by mixing a fraction of exact exchange in SCAN0, which mitigates the overestimated directional H-bonding strength. The softened water structure reduces the tendency of a proton within a water molecule to be donated to the neighboring water molecules, and therefore hinders the vibration of protons as described by the stretching mode. Consistently, the decompositions of the IR spectra show that the improved stretching band originates from both the inter- and intra-molecular contributions, which show a comparable blue shift compared to that from the SCAN functional. This is because the less polarized H-bond network by SCAN0 makes a proton less likely to be donated to a neighboring water molecule and therefore increases the covalent bond strength as evidenced by the shortened OH bond length as shown in Table~\ref{tbl_compare} or Fig.~\ref{fig:gr}. Therefore, it is not surprising that the intra- and inter-molecular contributions to the stretching mode both become stiffened by the exact exchange in SCAN0.

We next draw our attention to the bending band of the IR spectrum, which is centered $\sim$ 1200 $cm^{-1}$. Compared with the prediction by SCAN, the SCAN0 functional predicts an improved bending band. In particular, a significant enhancement can be identified in the intensity of the bending motion, which should be attributed to the improved H-bond network as well. Unlike the stretching band, the intra- and inter-molecular contributions to the bending mode are competing with each other. The intra- and inter-molecular dipoles have positive and negative correlations, respectively, which results in the opposite signs of the above two decompositions at roughly the same frequencies as shown in Fig.~\ref{fig:ir} (b). In crystalline ice, the intensity of the bending band is rather weak due to the strong inter-molecular dipole-dipole correlation while the contrary is true in liquid water \cite{mallamace_evidence_2007, imoto_molecular_2013}. Due to the artificially strengthened H-bond network described by the SCAN functional, the inter-molecular negative dipolar correlation is overestimated, which results in a weaker spectral intensity than that of experiment as shown in Fig.~\ref{fig:ir} (a). Whereas, the softened structure of liquid water predicted by SCAN0 suppresses the negative correlation of inter-molecular dipole moments and instead promotes the positive intra-molecular dipole correlation. The above effect gives an important correction to the bending band of the IR spectrum as seen by the increased overall intensity towards the experimental direction.

\begin{figure}
	\includegraphics[width=0.45\textwidth]{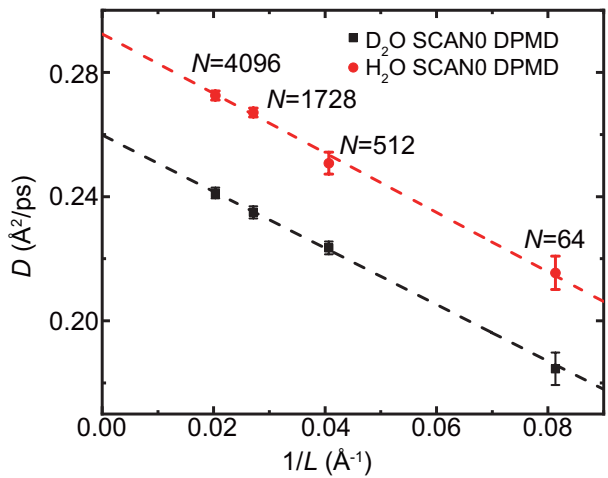}
	\caption{(a) Diffusion coefficients of liquid H$_2$O and D$_2$O from SCAN0 DPMD simulation at 330 K with different sizes of the simulation cell.}\label{fig:diffusion}
\end{figure}

Finally, we analyze spectral features in the low-frequency range, which are composed of the hindered translation band centered at $\sim$ 200 $cm^{-1}$ and the libration band center at $\sim$ 500 $cm^{-1}$. It can be seen that the IR spectrum predicted by SCAN0 also shows better agreement with experiment compared to SCAN. In particular, the improvement in the main peak of the libration band is more noticeable. The decompositions of the IR spectrum into its inter- and intra-molecular contributions, shows that the improvement originates from the intra-molecular part. The positive intra-molecular dipolar correlation is slightly enhanced under the weakened H-bond network by SCAN0, which hampers the water libration motion by shifting its band to a slightly higher frequency range.

Besides the improvement to the IR spectrum compared to SCAN, our SCAN0 DPMD simulation also captures the spectral difference between heavy and light water accurately as shown in Figs.~\ref{fig:ir} (a) and (c). Because of the lighter mass of H than D, all the four spectral features in light water are systematically higher than those in heavy water. A difference can be identified in the stretching/bending band between light water at 3475/1667 $cm^{-1}$ and heavy water at 2507/1219 $cm^{-1}$, respectively, which originates from molecular vibration in the vapor phase. In the low frequencies range, the spectral differences are relatively smaller. The translation/libration band is located at 208/675 $cm^{-1}$ for light water and 227/486 $cm^{-1}$ for heavy water. Not surprisingly, the reduced isotope effects are due to the fact that collective vibrations become more important in the lower frequency region, which is less affected by the mass difference between H and D.

\subsubsection{Diffusivity}

The self-diffusion coefficient is an important physical quantity that differentiates liquid water from crystalline ice. It has been well recognized that the calculation of the diffusion coefficient of water is a challenge for AIMD. The difficulty lies in the fact that the predicted diffusivity of water, which is modeled by finite box size under periodic boundary conditions, must be corrected before a meaningful comparison can be made with the experimental value $D_{\infty}$, which corresponds to the diffusivity of water in its thermodynamic limit. Based on the Kirkwood-Riseman theory of polymer diffusion, the correction term adopts the form of $D_{\infty}=D(L)+\frac{k_{B}T\zeta}{6\pi\eta L}$ \cite{04JPCB-Yeh}, where $D(L)$ is the diffusivity obtained in a system with cell length $L$, $\zeta$ is a numerical coefficient of 2.837, and $\eta$ is the shear viscosity. In the above, $D(L)$ approaches $D_{\infty}$ rather slowly with a $1/L$ scaling. In practice, $D_{\infty}$ is obtained by linear interpolation. Therefore, the accurate prediction demands water models of large box size containing thousands of molecules, which is computational challenging for DFT-based AIMD. In this work, utilizing the SCAN0 deep potential model, a series of simulations with cell sizes ranging from 64 to 4096 water molecules were conducted for both light and heavy water. For each cell size, four independent simulations were conducted for 300 ps. The diffusion coefficients were calculated by linear fittings of the mean square displacement of the molecular center of mass in the time interval of 1$\sim$10 ps. The error bars were estimated as the root mean square error of the four diffusion coefficients obtained from the four independent trajectories. As shown in Fig.~\ref{fig:diffusion}, the extrapolated values of D to infinite system size are 0.26 and 0.29 \r{A}$^2/$ps for D$_2$O and H$_2$O, respectively, which is in qualitative agreement with the room temperature experimental result of 0.20 and 0.24 \r{A}$^2/$ps for D$_2$O and H$_2$O \cite{73JPC-Mills}. It can be noticed that the diffusion coefficient 0.29 \r{A}$^2/$ps of light water predicted by the SCAN0 DPMD simulation at 330 K is slightly larger than the 0.26 \r{A}$^2/$ps predicted by the SCAN neural network \cite{yao_temperature_2020} at 330 K. This is because the more weakly H-bonded water molecules described by the SCAN0 than that by the SCAN functional promotes the diffusion of water.

The diffusion coefficient displays the translational dynamics of the H-bonds, while the rotational correlation time reflects the rotational dynamics which can be calculated by the integration of the second order orientational correlation function as:$$\tau_2=\int_0^{\infty}\langle{P_2}[\textbf{\emph{u}}(0){\cdot}\textbf{\emph{u}}(t)]\rangle{dt},$$
where $P_2$ is the second Legendre polynomial and $\textbf{\emph{u}}$ is the unit vector along the molecular vector of interest. In this work, $\textbf{\emph{u}}$ is chosen to be the unit vector along the OH covalent bonds of the water molecules. The values of $\tau_2$ for heavy and light water obtained from the SCAN0 DPMD simulation are 1.9 and 1.7 ps, respectively, which is in qualitative agreement with the nuclear magnetic resonance experimental values of 2.4 ps \cite{ropp_rotational_nodate} for heavy water and 1.7$\sim$2.6 ps \cite{lankhorst_determination_1982, smith_proton_1966} for light water. Compared to the $\tau_2$ of 3.3 ps for light water predicted by the SCAN neural network \cite{yao_temperature_2020} at 330 K, the $\tau_2$ of 1.7 ps predicted by our SCAN0 DPMD simulation indicates a faster rotational dynamics, which is in accordance with the faster translations dynamics by SCAN0 and can also be explained by the weaker hydrogen bonds due to the mitigation of self-interaction error.

\subsection{Electronic properties}

\subsubsection{Electronic DOS and bandgaps}

\begin{figure}
	\includegraphics[width=0.425\textwidth]{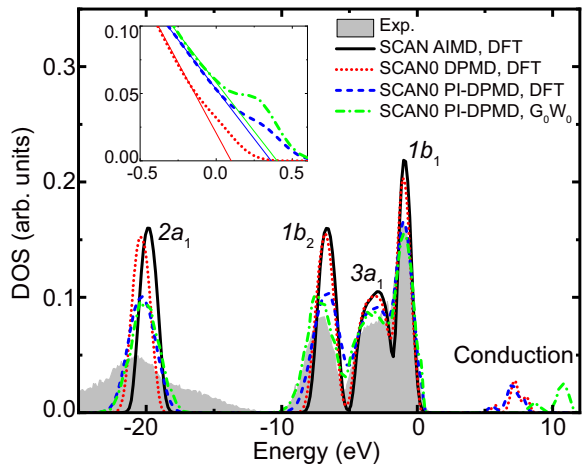}
	\caption{(a) Electronic DOS of liquid water calculated using the DFT and G$_0$W$_0$ methods based on the configurations predicted by SCAN AIMD \cite{17PNAS-Chen}, SCAN0 DPMD, and SCAN0 PI-DPMD at 330 K as well as the experimental photoemission spectra \cite{winter_full_2004} at room temperature. The $1b_1$ peak position is used to align all calculations. The inset shows the linearly extrapolated threshold of the valence band maximum.}\label{fig:dos}
\end{figure}

The electronic structure of liquid water can be efficiently probed by spectroscopy experiments that involving single-particle excitations. The valence band and conduction band can be extracted from the experimental data obtained by direct and indirect PES measurements  \cite{bernas_electronic_1997}, respectively. Moreover, the energy difference between the conduction band minimum and valence band maximum gives the quasiparticle (QP) bandgap. Physically, PES measures the energy of adding or removing a single-particle excitation from the system, which can be understood as electron or hole quasiparticles dressed by interactions with the system. In practice, these quasiparticles can be described by Hedin's GW approximation \cite{hedin_new_1965} within many-body perturbation theory. Within the GW approximation, the QP energies are typically calculated as one-shot correction without self-consistency (i.e., G$_0$W$_0$) on top of a DFT mean field, making electronic structure calculations within GW an important benchmark for the accuracy of the DFT electronic ground state.

In Fig.~\ref{fig:dos}, we present the DOS predicted by both DFT and G$_0$W$_0$ calculations, with the molecular configurations obtained from SCAN AIMD \cite{17PNAS-Chen}, SCAN0 DPMD and SCAN0 PI-DPMD simulations. The experimental PES result \cite{winter_full_2004} is also presented for comparison. In Fig.~\ref{fig:dos}, the four peaks with energies smaller than zero belong to the valence band and can be assigned to the $2a_1$, $1b_2$, $3a_1$, and $1b_1$ orbitals based on the spatial symmetries of a water molecule, while the peaks having energies above the Fermi level belong to the conduction bands. The DFT results in Fig.~\ref{fig:dos} show that, as the functional changes from SCAN to SCAN0, the energy of the $2a_1$ orbital with respect to the $1b_1$ orbital changes from -18.9 to -19.4 eV, the latter of which is closer to the experimental result of -19.7 eV. Therefore, SCAN0 provides a more precise description of the strongly bound $2a_1$ peak than SCAN. The above can be explained by the fact that the $2a_1$ orbital is primarily composed of the $2s$ electrons of the oxygen atom, which are corrected significantly when the self-interaction error is mitigated by the exact exchange in SCAN0. The decrease of the $2a_1$ energy explains the increased covalency by SCAN0, as the bonding pairs of electrons in a water molecule are primarily come from a combination of the $1b_2$ and $2a_1$ orbitals with a small proportion of the $3a_1$ orbital \cite{xu_isotope_2020}.
When NQEs are incorporated by the PI method, the distribution of the DOS is broadened and closer to the experimental result because NQEs enable the protons to explore more configuration space that is inaccessible in classical simulations. The broadening of the $2a_1$ and $1b_2$ orbitals are more obvious than the broadening of the $3a_1$ and $1b_1$ orbitals because the $2a_1$ and $1b_2$ orbitals are highly correlated with the bonding pairs of electrons and the hydrogen atoms in water molecules are delocalized significantly under NQEs.

For the SCAN0 PI-DPMD configurations, when the calculation method of the electronic structure changes from DFT to G$_0$W$_0$, the DOS associated with the valence states barely changes (the small differences between theses two results are within the margin of error associated with the statistical fluctuations, as G$_0$W$_0$ results are only averaged on eight snapshots), but the bandgap at the G$_0$W$_0$ level is significantly improves compared to DFT.
While Kohn-Sham DFT is a formally exact theory for the ground-state total energy and density of a material, a Kohn-Sham electronic structure calculation at fixed electron number fails to capture the contribution to the fundamental band gap from the discontinuity of the exact exchange-correlation potential under a change of electron number \cite{perdew_density-functional_1982, perdew_physical_1983}. As a result, it is well-known that the bandgaps in insulating systems are underestimated \cite{cohen_insights_2008, swartz_ab_2013}. The underestimations are noticeable in DFT calculations by LDA and GGA functionals and become less severe when the nonlocal XC effects are considered in generalized Kohn-Sham such as hybrid functionals with exact exchange \cite{perdew_understanding_2017}. The above is well illustrated by the comparison of the bandgap of 4.9 eV predicted by the meta-GGA SCAN functional \cite{17PNAS-Chen} with the bandgap of 5.7 eV predicted by the hybrid SCAN0 functional, the latter of which is closer to the experimental result of 8.7$\pm0.6$ eV \cite{bernas_electronic_1997}. It needs to be mentioned that the above bandgaps of 4.9 \cite{17PNAS-Chen} and 5.7 eV are obtained by simply taking the energy difference between the highest occupied orbital and lowest unoccupied orbital. In practice, a better way to calculate the bandgap is determining the position of the valence band maximum and conduction band minimum as the linearly extrapolated threshold as shown in the inset of Fig.~\ref{fig:dos}, which leads to faster convergence of the valence band maximum and conduction band minimum values with respect to cell sizes \cite{ambrosio_redox_2015, 16L-Chen}.
In this way, the bandgaps predicted by DFT based on the SCAN0 DPMD and PI-DPMD configurations are 5.1 and 4.6 eV, respectively. The quantum simulation reduceds the bandgap, in agreement with previous findings \cite{16L-Chen}. The G$_0$W$_0$ calculation corrects the bandgap from 4.6 to 7.5 eV, which is close to the experimental result of 8.7$\pm0.6$ eV \cite{bernas_electronic_1997}. Further improvement in the bandgap can be achieved by the quasiparticle self-consistent GW method with vertex corrections \cite{16L-Chen}. However, that is beyond the scope of this work.

\subsubsection{Dielectric constant}

\begin{figure}
	\includegraphics[width=0.425\textwidth]{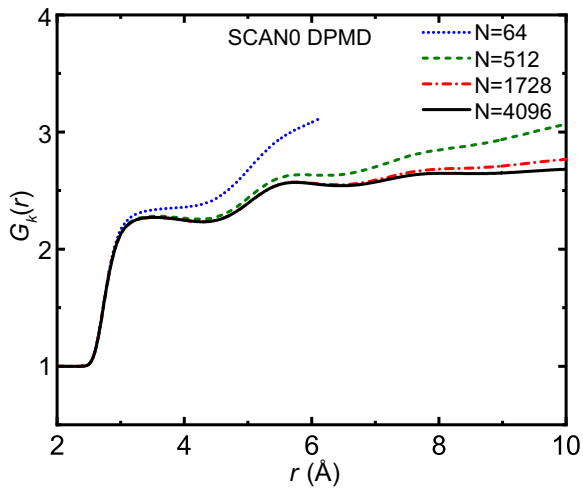}
	\caption{$G_k(r)$ of liquid water predicted by SCAN DPMD simulations with different cell sizes.}\label{fig:gk}
\end{figure}

The large dielectric constant is another anomalous property of liquid water, whose value is significantly larger than other materials with comparable electric dipoles  \cite{vidulich_dielectric_nodate}. A microscopic insight into the dielectric constant of liquid water is essential for understanding its properties as a solvent. However, the \emph{ab initio} calculation of the dielectric constant, $\epsilon_0$, of liquid water has long been a challenging prediction. Under the periodic boundary condition, the dielectric constant is connected with the fluctuations of electric dipole moments of water molecules and their correlations. Based on linear response theory, the dielectric constant of water can be calculated by the following equation \cite{07L-Sharma, adams_theory_1981, neumann_dipole_1983}:
\begin{equation}\label{eq:epsilon}
\epsilon_0=1+\frac{4\pi}{3}\frac{\rho\mu^2G_k}{k_{B}T},
\end{equation}
where $\mu$ is the average molecular dipole moment of the system, $\rho$ is the molecular number density, $k_B$ is the Boltzmann's constant, and $T$ is the temperature. $G_k$ is the large-$r$ limit of the correlation factor $G_k(r)$, which contains the integration of correlations among the dipole moments of different molecules and can be calculated as \cite{07L-Sharma}: $$G_k(r)=1+\frac{{\int}d\textbf{r}{\langle}\sum_{i{\neq}j}\bm{\mu}_i\cdot\bm{\mu}_j\delta(\textbf{\emph{r}}+\textbf{\emph{r}}_i-\textbf{\emph{r}}_j){\rangle}}{N{\langle}\mu^2{\rangle}}.$$
$G_k(r) = 1$ indicates there is no correlation between the dipoles of different molecules. In liquid water, the dipoles are highly correlated due to the H-bond network. The resulting $G_k(r)$ is significantly larger than 1, which gives rise to the large dielectric constant. In the above, the dipole moment can be rigorously derived from ion positions and Wannier centers that representing electron positions. Previous studies have shown that $\epsilon_0$ can be qualitatively computed by a 64-molecule cell \cite{07L-Sharma}, but large error bars remain due to a relatively small cell and a short trajectory. The difficulty lies in the fact that the correlation function converges rather slowly with $r$ for correlated dipoles as shown in Fig.~\ref{fig:gk} (a). Utilizing the SCAN0 DPMD and PI-DPMD simulation, the dielectric constants of water were calculated with simulation cells containing 64 to 4096 molecules at 330 K. Each SCAN0 classical simulation lasted for 2 ns and each SCAN0 quantum simulation lasted for 500 ps. It can be seen that $\epsilon_0$ converges slowly at the correlation distance r=10 \r{A} when the 4096-molecule cell is adopted. The SCAN0 classical and quantum simulation yield a dielectric constant of 76.06 and 83.57, respectively, which is close to the experimental result of 78.39 \cite{vidulich_dielectric_nodate}. As shown by the factor $\mu^2G_k(r)$ in Eq.~\ref{eq:epsilon}, the large dielectric constant of liquid water is contributed by both the molecular dipole moment and $G_k(r)$. The value of $G_k(r)$ predicted by SCAN0 classical and quantum simulations are 2.68 and 2.67, respectively. While the value of $\mu$ obtained from SCAN0 classical and quantum simulations are 2.97 and 3.11 Deybe, respectively.

\section{CONCLUSION}

In conclusion, we have investigated liquid water based on DFT, in which the hybrid meta-GGA functional, SCAN0, was employed to approximate the exchange-correlation effect among electrons. The molecular structure of the liquid, the dynamics of the H-bond network, and the electronic properties of water were predicted and compared to available experimental measurements. In particular, we applied deep neural networks as implemented in the DeePMD and Deep Wannier methods, which were trained on the potential energy surface and electronic structure obtained for the electronic ground state described by SCAN0-DFT. The applications of the deep neural network potentials enable us to model water in much larger simulation boxes with significantly longer simulation time than those in regular DFT, which thereby enabled prediction of the macroscopic water properties by approaching the thermodynamic limit.

In addition to the existing 17 constraints satisfied by the SCAN functional, the nonlocal exact exchange mixed in SCAN0 further alleviates the self-interaction error. The reduced molecular polarizability using SCAN0 softens the H-bond network of water towards the experimental direction as evidenced by the comparison of predicted RDFs with those obtained in scattering experiments. Since SCAN0 water is more loosely bonded by the H-bond network, the water molecules are more diffusive and less correlated in spatial and temporal dimensions. The improvements in water dynamics have been demonstrated in IR spectra and diffusivity by comparison between theory and experiments. The spatial correlations, as represented by the dipole-dipole correlation in the dielectric response, were also accurately determined. Moreover, the predicted static dielectric constant is in good agreement with experiments. The electronic band structure and band gap of water were determined by the GW self-energy approach as implemented in the Berkeley GW code package, which agree well with the PES experimental measurements. In all predicted static properties, the effects of including NQEs were found to give rise to mainly a broadening effect due to the delocalized protons as widely reported in the literature.

Our studies present the state-of-the-art theoretical modeling of water on the fourth rung of Jacob's ladder. Although there is improved accuracy in predicted water properties, some important discrepancies remain. The fact that the predicted water structure at 330 K is in accurate agreement with experiment under ambient conditions indicates that the predicted H-bond network by SCAN0 is still slightly overstructured. The water density predicted by SCAN0 is improved over that by SCAN, but it is still a little overestimated. To address these remaining issues, more sophisticated treatment of the self-interaction correction and long-range van der Waals interactions should be included in DFT along with nonlocal exchange-correlation functional constructions at higher rungs of Jacob's ladder.

\begin{acknowledgments}

This work was supported by National Science Foundation through Award No. DMR-2053195 and No. DMR-1552287. This research used resources of the National Energy Research Scientific Computing Center (NERSC), which is supported by the U.S. Department of Energy (DOE), Office of Science under Contract No. DE-AC02-05CH11231. The work of F. T. and M. L. K. was supported by the Computational Chemical Center: Chemistry in Solution and at Interfaces funded by the DOE under Award No. DESC0019394.
The work of J. P. P. was supported by the U.S. National Science Foundation under Grant No. DMR-1939528, with a contribution from CTMC - Division of Chemistry. The work of D. Y. Q. was supported by the Center for Computational Study of Excited State Phenomena in Energy Materials (CSEPEM), which is funded by the U.S. Department of Energy, Office of Science, Basic Energy Sciences, Materials Sciences and Engineering Division under Contract No. DE-AC02-05CH11231.
This research includes calculations carried out on HPC resources supported in part by the National Science Foundation through major research instrumentation grant number 1625061 and by the U.S. Army Research Laboratory under contract No. W911NF-16-2-0189.
This research used resources of the Oak Ridge Leadership Computing Facility at the Oak Ridge National Laboratory, which is supported by the Office of Science of the U.S. Department of Energy under Contract No. DE-AC05-00OR22725.
\end{acknowledgments}

\bibliography{reference}
\end{document}